\documentclass[twocolumn,aps,prb,showpacs]{revtex4}
\usepackage{graphics,bm}
\usepackage{amssymb}
\usepackage{epsfig}
\usepackage{epsf}

\def\frac#1#2{{\textstyle{#1 \over #2}}}

\def\Re{{\rm Re}}
\def\Im{{\rm Im}}
\def\be{\begin{equation}} \def\ee{\end{equation}}
\def\bea{\begin{eqnarray}} \def\eea{\end{eqnarray}}

\def\nn{\nonumber}

\def\vk{{\vec k}}
\def\vkp{{\vec k}'}

\def\vq{{\vec q}}
\def\vp{{\vec p}}

\begin{document}

\title{Characterizing Featureless Mott Insulating State by Quasiparticle Interferences - A Dynamical Mean Field Theory Prospect}
\author{Shantanu Mukherjee}
\author{Wei-Cheng Lee}
\email{wlee@binghamton.edu}
\affiliation{Department of Physics, Applied Physics, and Astronomy, Binghamton University - State University of New York, Binghamton, USA}

\date{\today}

\begin{abstract}
The quasiparticle interferences (QPIs) of the featureless Mott insulators are investigated by a $T$-matrix formalism implemented with the dynamical mean-field theory ($T$-DMFT). 
In the Mott insulating state, due to the singularity at zero frequency in the real part of the electron self energy ($\Re \Sigma(\omega) \sim \eta/\omega$) predicted by DMFT, 
where $\eta$ can be considered as the \lq order parameter\rq~for the Mott insulating state,
QPIs are completely washed out at small bias voltages. However, a further analysis shows that 
$\Re \Sigma(\omega)$ serves as an energy-dependent chemical potential shift. As a result,
the effective bias voltage seen by the system is  $eV' = eV - \Re \Sigma(eV)$, which leads to a critical bias voltage 
$eV_c \sim \sqrt{\eta}$ satisfying $eV'=0$ if and only if $\eta$ is non-zero. Consequently, the same QPI patterns produced by the non-interacting Fermi surfaces appears at this 
critical bias voltage $eV_c$ in the Mott insulating state. We propose that this re-entry of non-interacting QPI patterns at $eV_c$ could serve as an experimental signature of the Mott insulating
state, and the 'order parameter' can be experimentally measured as $\eta \sim (eV_c)^2$.
\end{abstract}


\maketitle

{\it Introduction} --
Mottness, the physics of understanding how the insulating state arises from a partially-filled conduction band due to strong local interactions, has been one of the most challenging subjects in condensed matter physics. It could potentially hold the key to understanding the mechanism of high-temperature superconductivity observed in various materials with a narrow bandwidth\cite{mott1,mott2,mott3}.
The essential physics of Mottness can emerge from the single orbital Hubbard model,
\bea
H_{Hubbard} &=& H_t + H_U\nn\\
H_t &=& -t \sum_{<i,j>,\sigma} c^\dagger_{i\sigma} c_{j\sigma} + h.c.,\nn\\
H_U &=& U\sum_i n_{i\uparrow}n_{i\downarrow},
\label{hubbard}
\eea
where $t$ is the hopping parameter between nearest neighbor sites $<i,j>$, and $U$ is the on-site Coulomb interaction.  
Although we analyze a single orbital model here, our conclusions are also applicable to multi-orbitals systems.
While the Hubbard model
has been successful in conceptually demonstrating both the metallic and the insulating states in two extreme limits of $U/t << 1$ and $U/t >> 1$, the nature of the transition 
from the metallic to the insulating states as a function of $U/t$ is still object of active investigation. 
One particular difficulty is that the Mott transition does not seem to involve breaking 
of symmetry, and consequently such a transition can not be studied within the framework of Landau-Ginzburg theory.
The change in the temperature dependence of dc resitivity\cite{mott1} is often used as an indicator of the Mott transition, but it depends on many 
other details of the materials, e.g., impurity and disorder. As a result, the dc resitivity can not be a conclusive experimental signature for the Mott transition.
Other possible candidates for an \lq order parameter\rq~of the Mott transition, such as, the density of states at Fermi energy\cite{brinkman1970,fresard1992} or the number of double occupancy\cite{castellani1979}, have been discussed in the literature. However, these physical quantities are also not ideal candidates for an order parameter behavior since for example, the density of states at Fermi energy is zero in the insulating state and therefore does not contain information about the strength of Mott insulator. In other words, an order parameter for the Mott transition that is theoretically well-defined and experimentally observable is lacking. 

The dynamical mean-field theory (DMFT) has been shown to be a sophisticated theoretical approach to 
understand the Mott insulating state\cite{dmft1,dmft2,tremblay2010,fiete2014,dmft3}.
DMFT maps the problem of solving the Hubbard model to another one of solving the Anderson impurity model which is much better understood. Such a mapping is shown to be \lq exact\rq~in the limit of 
spatial dimension $d$ going to infinity. The price paid for this mapping, however, is that the momentum-dependence of the electron self energy is completely neglected, but the resulting 
electron self energy $\Sigma(\omega)$ includes {\it all} the local quantum fluctuations. Since the Mott insulating state concerns mostly the local physics, $\Sigma(\omega)$ from DMFT captures a lot of 
information that can not be accessed by other approaches easily, despite its lacking of the momentum dependence.

In DMFT for the case of half-filling\cite{dmft1}, the real part of the electron self energy has the following singular behavior at low frequency\cite{supp}
\be
\Re \Sigma(\omega) = \frac{\eta}{\omega} + O(\omega),
\label{order}
\ee
Here $\eta$ is given by,
\be
\eta = \big[-\frac{1}{\pi}\int_{-\infty}^{\infty} \frac{d\epsilon}{\epsilon^2} \sum_{\vk} \Im G_b(\vk,\epsilon+i0^+)\big]^{-1}, 
\ee
where
$G_b(\vk,\omega)$ is the Green function obtained from DMFT with the following form,
\be
G_b(\vk,\omega)= \big[\omega-E(\vk) - \Sigma(\omega)\big]^{-1},
\label{gb}
\ee
$E(\vk) = \epsilon(\vk)-\mu$, where $\epsilon(\vk) = -2t(\cos k_x + \cos k_y)$, and $\mu$ is the chemical potential. The constant term $-U/2$ has been absorbed into the chemical potential.

This $1/\omega$ singularity in $\Re \Sigma(\omega)$ is inherited from the self energy in the atomic limit of the Hubbard model ($t=0$), and DMFT captures it well even with non-zero $t$.
It can be shown\cite{dmft1} that $\eta = 0$ in the metallic phase while $\eta\neq 0$ in the Mott insulating phase. 
Moreover, a larger $\eta$ corresponds to a stronger Mott insulating state. 
Due to the above reasons, $\eta$ has been proposed to have properties similar to an order parameter of the Mott insulating state\cite{kotliar2000,dmft1}.
However, $\eta$ remains a purely theoretically-defined \lq order parameter\rq~since no experimental probe has been applied to measure it.

In this paper, we propose that the quasiparticle interferences (QPI's) from impurities on a material surfaces which are measured by spectroscopic imaging scanning tunelling microscopy (SI-STM) could be a feasible tool to 
directly measure $\eta$. 
QPIs have been widely used to extract the spectroscopic information of the electronic structure in materials including, but not limited to, cuprates\cite{qpicupratereview,qpicuprate,pegorcdw}, 
iron-based superconductors\cite{qpife1,qpife2,qpife3}, 
topological insulators\cite{qpiti1,qpiti2,leewcqpiti,huqpiti}, Sr$_3$Ru$_2$O$_7$\cite{qpi327,leewcqpi327,leewcqpi3272}, and heavy fermion systems\cite{pegor1,pegor2}.
On the theoretical side, the $T$-matrix formalism\cite{wanglee2003,balatsky2006,leewcqpi327,leewcqpi3272} has been shown to be a well-established method to compute the QPI images.
We develop a $T$-matrix formalism implemented with DMFT ($T$-DMFT) to analyze the QPI patterns in the Mott insulating state. 
We find that at small bias voltage, no meaningful QPI images are found due to the
singular behavior of the electron self energy given in Eq. \ref{order}, as expected. However, by analyzing the $T$-DMFT formalism, we further find that the QPI patterns resembling the one produced 
by the non-interacting Fermi surface appear at a non-zero critical bias voltage $V_c$, and this critical bias voltage is directly related to $\eta$, the \lq order parameter\rq~of the Mott insulating state 
given in Eq. \ref{order}. We propose that this novel re-entry of non-interacting QPI at high bias voltage could serve as an experimental signature of the Mott transition, and
the order parameter can be obtained by $\eta \sim (eV_c)^2$ accordingly.

{\it Formalism} -- 
We start from the following model Hamiltonian of
\bea
H&=&H_{Hubbard} + H_{imp},\nn\\
H_{imp} &=& \sum_{\vk,\vkp} \sum_\sigma V_{\vk,\vkp} c^\dagger_{\vk\sigma} c_{\vkp\sigma} ,
\eea
where $H_{Hubbard}$ is the Hubbard model described in Eq. \ref{hubbard}, and $V_{\vk,\vkp}$ is the impurity scattering matrix element which is assumed to be 
spin-independent.
To specifically compute the QPI image, we employ a $T$-matrix
approach \cite{wanglee2003,balatsky2006} implemented with DMFT, and the procedure is given below.

Following Ref. [\onlinecite{dmft1}],
we first solve $H_{Hubbard}$ with DMFT using Hirsch-Fye quantum Monte Carlo algorithm as the impurity solver. The imaginary time interval $[0,\beta]$ is divided into 
$L=128$ slices, and the maximum Matsubara frequency used is $n_{max} = 8192$. 
We use the semicircular density of states with half-bandwidth $D$. After the self-consistent calculation is converged, the time-ordered electron self energy in 
Matsubara frequency $\Sigma(i\omega_n)$ can be evaulated directly. 
In order to obtain the electron Green function in real frequency $G_b(\vk,\omega)$, 
a numerical calculation of the analytic continuation on $\Sigma(i\omega_n)$ is necessary. 
We adopt the continuous-pole-expansion method\cite{staar2014} recently developed by Staar, {\it et al.} to numerically perform the analytic continuation on $\Sigma(i\omega_n)$. 
This method is shown to be an efficient approach to find an accurate and unambiguous result of the analytic continuation within a finite range of frequency, which is necessary for the current study.
Finally, the electron Green function with DMFT $G_b(\vk,\omega)$ given in Eq. \ref{gb} can be obtained after the analytic continuation on $\Sigma(i\omega_n)$ is performed.

Once $H_{Hubbard}$ is solved by DMFT, we treat impurity scatterings necessary for the QPI images as a perturbation around the DMFT solution. 
If the impurity scatterings are weak, $T$-matrix formalism \cite{wanglee2003,balatsky2006} allows us to sum over all the vertex corrections due to the impurity scatterings with 
$G_b(\vk,\omega)$ as the \lq bare\rq~Green function.
The full Green function including impurity scatterings can be expressed as\cite{wanglee2003,balatsky2006}
\bea
G(\vk,\vkp,\omega)&=&G_b(\vk,\omega)\,
\delta_{\vk,\vkp} + G_b(\vk,\omega)\,
T_{\vk,\vkp}(\omega) \, G_b(\vkp,\omega)\nn\\
\label{gtg} \eea
where the $T$-matrix satisfies
\bea 
T_{\vk,\vkp}(\omega)=V_{\vk,\vkp} +\int\!\!d^2p\>
V_{\vk,\vp} \, G_b(\vp,\omega) \,
T_{\vp,\vkp}(\omega)\ , \label{tmatrix} 
\eea
Typically, SI-STM is performed on a nicely-cleaved surface on which the impurities are usually apart from each other.
In this case, the STM conductances ($\frac{dI}{dV}$) at positions near a single impurity are measured, and consequently it is sufficient to include only one impurity 
in the calculation \cite{wanglee2003,balatsky2006}.
For the case of the single impurity, $V_{\vk,\vkp} = V_0$, and $T_{\vk,\vkp}(\omega)$ can be further reduced to $T(\omega)$. As a result, $T(\omega)$ can be evaluated from 
$G_b(\vk,\omega)$ by
\be
\frac{1}{T(\omega)} = \frac{1}{V_0} - \int\!\!d^2\vk\>
G_b(\vk,\omega)
\ee

It is well-established \cite{wanglee2003,balatsky2006} that $\frac{dI}{dV}(\vec{r},eV)$ at bias voltage $V$ is proportional to the
local density of states $\rho(\vec{r},eV)=\sum_\sigma\rho_\sigma(\vec{r},eV)$
where $\rho_\sigma(\vec{r},\omega)=\Im
G_\sigma(\vec{r},\vec{r},\omega)$. The QPI image in the $\vec{q}$ map 
, namely $\rho(\vec{q},\omega)$, is just the Fourier component of $\rho(\vec{r},eV)$.  After a straightforward calculation, the $T$-DMFT formalism leads to
\bea
&&\rho(\vq,eV) - \rho_0(eV) =\nn\\
&& \Im\big\{T(eV) \int \!\! d^2k\>G_b(\vk,eV)G_b(\vk+\vq,eV)\big\},
\label{rhoq}
\eea
where $\rho_0(eV)$ is the density of states without the impurity scatterings which will be dropped out from now on since it is just an unimportant constant.

\begin{figure}
\includegraphics[width=\columnwidth]{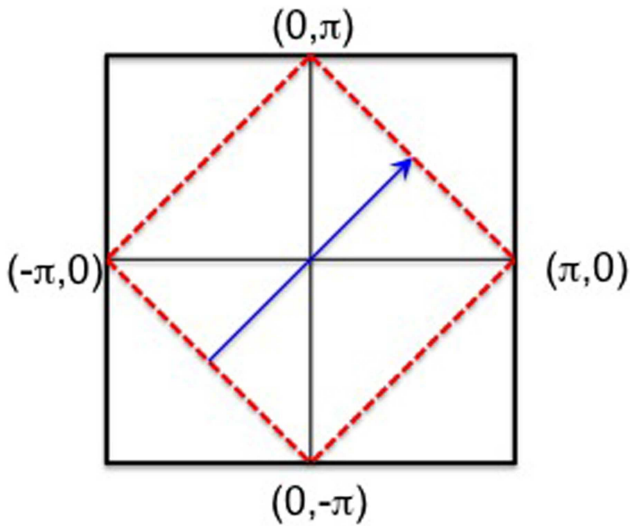}
\caption{\label{fig:nonintqpi} The Fermi surface of $H_{Hubbard}$ with $U=0$ at half-filling. The red dashed lines represent the Fermi surfaces, 
and the blue arrow indicates the momentum (in this case, $(\pi,\pi)$) connecting two points on the Fermi surface with large joint density of states. }
\end{figure}

{\it Results} --
For the non-interacting case, $\Sigma(\omega)$ in Eq. \ref{gb} is zero, and the peaks in the $\vq$ map of $\rho(\vq,eV)$ are corresponding to the momenta connecting two points 
with large joint density of states on the equal energy contour at energy $eV$ away from the Fermi energy \cite{wanglee2003,balatsky2006}.
For example, the Fermi surface of $H_{Hubbard}$ with $U=0$ at half-filling is depicted by the red dashed lines in Fig. \ref{fig:nonintqpi}. The $\vq$ map of $\rho(\vq,eV=0)$ has a 
peak at $\vec{q}=(\pi,\pi)$ corresponding to the blue arrow in Fig. \ref{fig:nonintqpi}.

As the interaction is present, 
\bea
G_b(\vk,\omega)&=& \big[\omega-E(\vk) - \Sigma(\omega)\big]^{-1}\nn\\
&=&  \big[\omega^\prime - E(\vk) - i\Im \Sigma(\omega)\big]^{-1},
\eea
where $\omega^\prime = \omega - \Re \Sigma(eV)$.
Consequently, Eq. \ref{rhoq} becomes
\bea
&&\rho(\vq,eV) = \Im\big\{T(eV)\\
&\times& \int \frac{d^2\vk}{[eV' - E(\vk) - i\Im \Sigma(eV)][eV' - E(\vk+\vq) - i\Im \Sigma(eV)]}\big\}.\nn
\label{rhoq2}
\eea
It is now clear that $eV' = eV - \Re \Sigma(eV)$ is the {\it effective} bias voltage shifted due to the present of $\Re \Sigma(eV)$. The QPI images at bias voltage $eV$ in this case is in fact the same 
as the non-interacting one at $eV'$ except the extra broadening resulted from $\Im \Sigma(eV)$.

\begin{figure}
\includegraphics[width=\columnwidth]{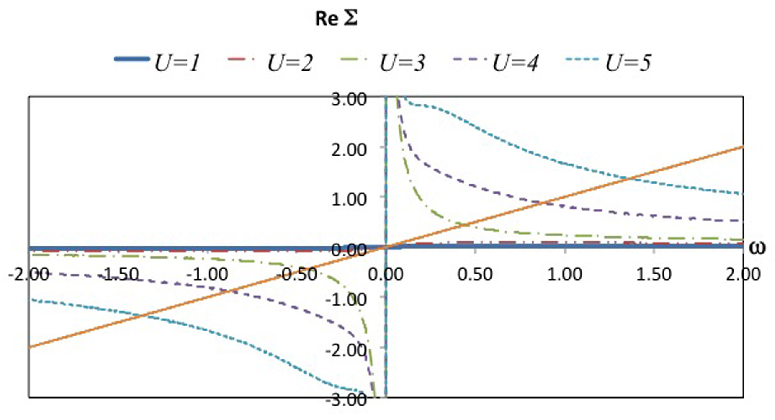}
\includegraphics[width=\columnwidth]{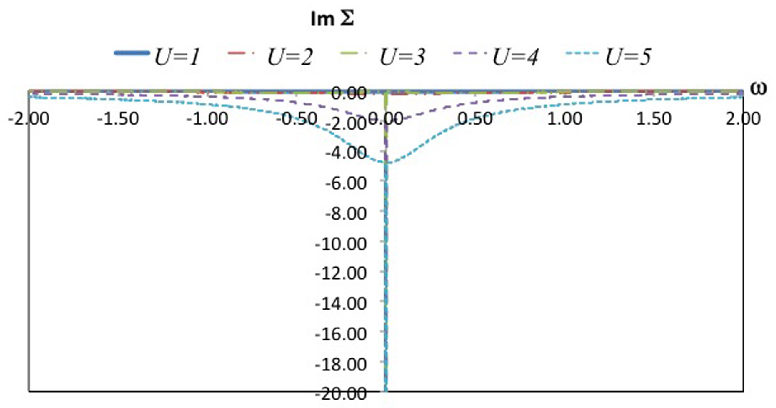}
\caption{\label{fig:self} The real (top) and imaginary (bottom) parts of the electron self energy from DMFT for $U=1,2,3,4,5$ with temperature $k_B T = 0.1$. Note that the unit of energies is 
$D$ as used in Ref. \onlinecite{dmft1}. The critical bias $V_c$ is the point 
crossed by the line of $\omega$ and $\Re \Sigma(\omega)$. For $U=1,2$, the system is not in the Mott insulating state, thus $V_c=0$.}
\end{figure}

The electron self energy $\Sigma(\omega)$ obtained from DMFT calculation for various valules of $U$ is plotted in Fig. \ref{fig:self}. Note that both $U$ and $k_B T$ are in the 
unit of $D$ as used in Ref. 
\onlinecite{dmft1}. For small $U$, the self energy is small, and therefore 
the QPIs are not expected to deviate from the ones obtained in the non-interacting case. The situation, however, changes dramatically as $U$ is large enough for the occurrence of the Mott
insulating state ($2<U_c<3$ in the current calculation). In the Mott insulating state, $\Im \Sigma(\omega)$ has a delta-function-like peak at $\omega=0$, while $\Re \Sigma(\omega)$ exhibits a 
divergence of $1/\omega$ at small frequency. These behaviors of self energy make $G_b(\vk,\omega)$ very small at small frequency, 
which washes away any meaningful features in the QPI images at small bias voltage.

Intriguing features appear at high bias voltage. The singular behavior of $\Re \Sigma(\omega)$ described in Eq. \ref{order} guarantees the existence of a 
critical bias voltage $eV_c$ satisfying 
\be
eV'_c = eV_c - \Re\Sigma(eV_c)= 0.
\ee
Adopting the expansion of $\Re\Sigma(\omega)$ given in Eq. \ref{order}, we find $eV_c \sim \sqrt{\eta}$. Although the precise value of $eV_c$ depends on the high order corrections, it is important 
to recognize that $eV_c$ scales with $\eta$, and $eV_c=0$ if and only if $\eta=0$.
In other words, {\it in the Mott insulating state, there is always a non-zero critical bias voltage $V_c$ at which the QPI 
pattern is exactly the same as the non-interacting one except extra broadening, and the \lq order parameter\rq~of the Mott insulating state can be indicated by $\eta \sim (eV_c)^2$.}

\begin{figure}
\includegraphics[width=\columnwidth]{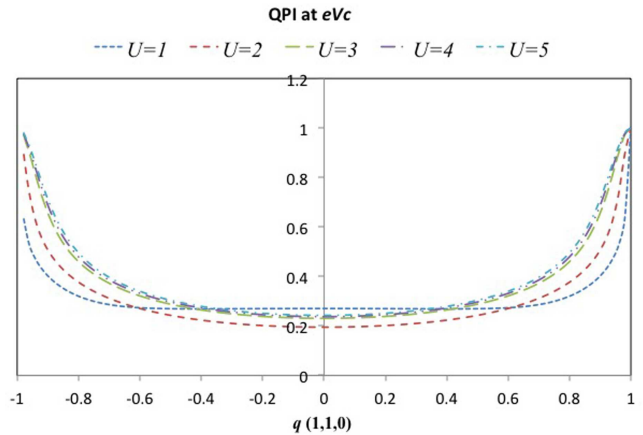}
\caption{\label{fig:qpi} The QPI at the critical bias voltage $V_c$ for different $U$ along the nodal direction. $\rho(\vq,eV_c)$ is normalized with respect to the peak value at $\vq=(\pi,\pi)$ for
each $U$. $eV_c=0,0,0.44,0.88,1.37$ for $U=1,2,3,4,5$.}
\end{figure}

To confirm this, we consider the one-band Hubbard model with the nearest neighbor hopping at half-filling. The non-interacting QPI exhibits a strong peak at $\vq = (\pi,\pi)$ at zero bias and the peak 
shifts to a smaller momentum as the bias voltage increases. Fig. \ref{fig:qpi} plots the QPI patterns along the nodal direction in the first Brillouin zone at the critical bias voltage $V_c$ 
for different $U$. All of them exhibit 
exactly the same peak at $\vq = (\pi,\pi)$, resembling the non-interacting QPI at zero bias. For $U=1,2$, the system is not in the Mott insulating state and $V_c=0$. 
For $U=3,4,5$, the non-interacting QPI pattern appears at $eV_c = 0.44, 0.88, 1.37$ respectively. Longer range hopping parameters can be added in order to capture more details in the Fermi surface, but these additions do not change the conclusions 
except the critical bias voltage might be slightly modified.

{\it Discussions and summary} --
The current result is primarily based on the feature that the electron self energy obtained from DMFT is momentum-independent. 
This feature is correct in the atomic limit of the Hubbard model\cite{dmft1}, but 
in general the electron self energy should depend on the momentum whenever $t/U$ is non-zero. Nevertheless, since we are mainly interested in the Mott insulating state in which $t/U$ is a 
small parameter, the momentum-dependence of the electron self energy should be very weak in the Mott insulating state, which has been discussed in previous 
studies\cite{chubukov1,chubukov2,phillips2010,leewc2011,dave2013,chubukov3}. As a result, the predicted re-entry of the non-interacting QPIs at a critical bias voltage should be robust even as the momentum dependence of the electron
self energy is considered.

It is worth being mentioned that the critical bias voltage is smaller than the Mott gap as we can see in Fig. \ref{fig:qpi}. The typical value of $D$ for cuprates is around 400 meV, which gives 
the critical bias voltage about 150 meV to 600 meV for $U=3-5$. The fundamental limitation on the energy resolution of SI-STM is the range of the energy within which the density of 
states of the STM tips remains flat. 
A typical STM tip has a flat density of states within $E_F\pm 500$ meV. As a result, although the critical voltage is a little higher than the typical range of the bias voltages where the SI-STM is 
routinely exploited, it is still measurable with the current techniques.

The existence of the critical bias voltage $eV_c$ is not limited to the electron self energy obtained from DMFT. In fact, any theory having an electron self energy with a diverging behavior at low 
frequency is likely to have a non-zero $eV_c$. From the Dyson's equation, the self energy can be expressed as $\Sigma = G_0^{-1}-G^{-1}$, where $G_0$ and $G$ are the non-interacting and the full 
Green functions. Since $G_0$ is non-zero always, $\Sigma$ could have a diverging behavior if and only if the full Green function has zeros at low frequency. The existence of zeros in full Green function 
has been recently found in a variety of theories for strongly correlated electrons\cite{zero2007,zero2012,zero2013,dave2013}, including the AdS-CFT holographic models\cite{ads1,ads2}. 
As a result, it is expected that the same result will be obtained if these theories could be incorporated into $T$-matrix formalism to compute QPIs. Further, note that although for computational simplicity we have focussed on the 2D Hubbard model which is applicable to a large class of materials including cuprates and Heavy-Fermions, our conclusions are more general and applicable to 3D systems as well.

In summary, we have employed the $T$-matrix formalism implemented with the dynamical mean-field theory (DMFT) to study the QPI patterns of the featureless Mott insulator. 
While QPIs at small bias voltages are completely washed out due to the 
singular electron self energy obtained from DMFT, we find that the QPI patterns resembling the non-interacting ones appear at a non-zero critical bias voltage. 
Since the existence of this non-zero critical bias voltage is a direct consequence of the singular behavior of $\Re \Sigma(\omega) \sim \eta/\omega$ in which $\eta$ could be thought of
the 'order parameter'  of the Mott insulating state, this novel re-entry of non-interacting QPI patterns at a finite critical bias voltage could serve as an experimental signature of the Mott 
transition, and the order parameter can be indicated by $\eta \sim (eV_c)^2$ accordingly.

{\it Acknowledgement} --
We are grateful for valuable discussions with P. Aynajian, M. J. Lawler, and P. W. Phillips.
W.-C. Lee thanks for the hospitality of KITP at UCSB while this manuscript was being initiated.
This work is mainly supported by a start up fund from Binghamton University and in part by the NSF under Grant No. NSF PHY11-25915 for the KITP program
of 'Magnetism, Bad Metals and Superconductivity: Iron Pnictides and Beyond'.


\end{document}